# Solid solution decomposition and Guinier-Preston zone formation in Al-Cu alloys: A kinetic theory with anisotropic interactions

A. Yu. Stroev[1,2], O. I. Gorbatov[3,4,5], Yu. N. Gornostyrev[4,6,7], P. A. Korzhavyi[3]

[1]*National Research Centre "Kurchatov Institute", 123182, Moscow, Russia*
[2]*Moscow Institute of Physics and Technology (State University), 141700, Dolgoprudny, Moscow region, Russia*
[3]*KTH Royal Institute of Technology, SE-100 44, Stockholm, Sweden*
[4]*Institute of Quantum Materials Science, 620072, Ekaterinburg, Russia*
[5]*Nosov Magnitogorsk State Technical University, 455000, Magnitogorsk, Russia*
[6]*Institute of Metal Physics, Ural Division RAS, 620219, Ekaterinburg, Russia*
[7]*Ural Federal University, 620002, Ekaterinburg, Russia*

Using methods of statistical kinetic theory parametrized with first-principles interatomic interactions that include chemical and strain contributions, we investigated the kinetics of decomposition and microstructure formation in Al-Cu alloys as a function of temperature and alloy concentration. We show that the decomposition of the solid solution forming platelets of copper, known as Guinier-Preston (GP) zones, includes several stages and that the transition from GP1 to GP2 zones is determined mainly by kinetic factors. With increasing temperature, the model predicts a gradual transition from platelet-like precipitates to equiaxial ones and at intermediate temperatures both precipitate morphologies may coexist.

## 1. INTRODUCTION

Al-based alloys strengthened by Cu precipitates are among the main construction materials when a combination of high strength and low specific weight is required [1], such as in the fuselage of an aircraft. Their high-strength characteristics are mainly due to the coherent metastable precipitates of Cu-rich particles that form after solid solution quenching and subsequent ageing or low-temperature annealing. In binary Al-Cu alloys such precipitates are thin platelets with {001} orientation, the so-called Guinier-Preston zones (GPZ) [2,3].

The nature of GPZ in Al alloys has been the subject of numerous investigations (see review articles [4,5,6]). It has been established that an increase in temperature or holding time results in the growth of GPZ and their transformation, which includes several stages: GP1 → θ″ (GP2) → θ′ → θ ($Al_2Cu$) phase. However, even today little is known about the first stages of GPZ formation. In particular, there is no clear picture of the process of initial agglomeration of Cu solutes to form the first clusters, and the role of vacancies and other lattice defects in the formation of the clusters remains unclear [7].

It is commonly believed that the elastic strains caused by the atomic size mismatch between solvent and solute elements plays a crucial role at the early stages of decomposition of an Al-Cu solid solution [8,9,10,11,12,13]. To describe the formation of GPZ a model has been proposed [9] that uses a cluster expansion of the alloy configurational energy at different length scales (mixed-space cluster expansion (MSCE)) and includes the solute-solute interaction energy as well as the coherency strain energy due to the size misfit between the alloy constituents.

In the MSCE approach, the GPZ formation is fully determined by the coherency strain energy [10,11]. This conclusion coincides with the results of the classical model of GPZ based on linear elasticity theory [8], but leaves open the question about precipitate nucleation which is beyond the scopes of linear elasticity. At the same time, such early stages of solid solution decomposition should be amenable to the traditional cluster expansion treatment, provided that the strain contribution is also expanded in terms of pair and many-body interactions. Such an approach was recently realized in Ref. [14] where GPZ formation was investigated using first-principles derived cluster interactions that included a chemical contribution and a strain contribution due to the relaxation of the lattice around the Cu solutes and their clusters. Planar clusters and platelets of Cu atoms with the {100} orientation were shown to be energetically favored in Al-Cu alloys. In subsequent Metropolis Monte Carlo simulations the temperature–concentration domain of GPZ formation was established, in agreement with experimental observations. However, the approach of Ref. [14] did not allow the authors to investigate neither the kinetics of GPZ formation nor the subsequent stages of their evolution with increasing temperature and/or holding time.

In the present work, we report studies on the kinetics of solid solution decomposition in the Al-Cu system, using statistical kinetic theory of alloys [15,16,17] in conjunction with first-principles parametrization of the effective interactions [14], taking into account both chemical and strain contributions. We show that kinetic factors may play a significant role at the early stages of solid solution decomposition and GPZ formation. The formalism of the model is introduced in Section 2. The modeling results are presented and discussed in Sections 3 and 4.

## 2. MODEL AND MASTER EQUATION FOR DECOMPOSITION KINETICS

The statistical alloy theory based on the master equation (ME) approach was proposed in Refs. [15,16,17]. Despite the simplification used (mean-field approximation, direct exchange mechanism of diffusion) it allows the detailed description of various stages of both the decomposition and microstructure formation in alloys.

The statistical theory uses the representation of the energy of an alloy $A_{1-c}B_c$ in the form of a cluster expansion (CE) [18]:

$$H = \frac{1}{2}\sum_{i,j} v_{ij}^{(2)} n_i n_j + \frac{1}{3!}\sum_{i,j,k} v_{ijk}^{(3)} n_i n_j n_k + \frac{1}{4!}\sum_{i,j,k,l} v_{ijkl}^{(4)} n_i n_j n_k n_l + ... \quad (1)$$

where $v_{ij}^{(2)}$, $v_{ijk}^{(3)}$, $v_{ijkl}^{(4)}$ are the energies of pair, triple, and quadruple effective cluster interactions of atoms of species B, $n_i$ the occupation number equal to 1 if site $i$ is occupied by an atom B or 0 otherwise. The microstructure evolution of the alloy is described by a time-dependent probability $P(\xi)$ of realization of a given set of occupation numbers $\xi \equiv \{n_i\}$, which obeys the master equation

$$\frac{dP(\xi)}{dt} = \sum_{\eta} (W(\xi,\eta) P(\eta) - W(\eta,\xi) P(\xi)), \quad (2)$$

where $W(\xi,\eta)$ is the probability of a $\xi \to \eta$ transition per unit of time, for which a conventional thermally activated atomic exchange model is used:

$$W_{ij} = n_i (1 - n_j) w^{eff} \exp\frac{1}{T} (\hat{E}^{in} - \hat{E}^{SP}). \quad (3)$$

(The applicability of such an approach is discussed in Ref. [16]). Here $T$ is the temperature (in energy units), $\hat{E}^{SP}$ the energy corresponding to the saddle point along the path of the diffusion jump of an atom from the initial to the final position, $\hat{E}^{in}$ the configurational energy for the initial position of the atom (before the jump), $w^{eff}$ the effective attempt frequency including the entropy contribution. As has been shown in Ref. [19] the realistic vacancy-mediated exchange mechanism can be described in terms of the equivalent direct exchange model used here. This approximation leads mainly to rescaling of the time but does not change the scenario of microstructure evolution nor the precipitate morphology. Similar simplification is often used in kinetic Monte Carlo (KMC) simulations (see, for instance, Ref. [20]) because it significantly speeds up the computation.

The probability $P(\xi)$ given by Eq. (2) may be expressed in a quasi-equilibrium form [16] as

$$P\{n_i\} = \exp\frac{1}{T} (\Omega + \sum_i \lambda_i n_i - H). \quad (4)$$

Here the parameters $\lambda_i$, which, in general, are time- and position-dependent, may be considered as on-site (local) chemical potentials, while the constant $\Omega = \Omega(\lambda_i)$ ensures normalization of the probability $P(\xi)$ to unity.

By multiplying Eq. (2) by $n_i$ and taking a sum over all configurations corresponding to the given macroscopic state [16], one obtains the quasi-equilibrium kinetic equation (QKE) [16]:

$$\frac{dc_i}{dt} = \sum_j 2 M_{ij} \sinh\frac{\lambda_j - \lambda_i}{2T} \quad (5)$$

Here the average site occupancy is

$$c_i = \bar{n}_i = \sum_{\{n_j\}} n_i P\{n_j\}, \quad (6)$$

and $M_{ij}$ the generalized mobility. In the mean-field approximation (MFA) $\lambda_i$ and $M_{ij}$ take the form [16]:

$$\lambda_i = T \ln\frac{c_i}{1-c_i} + \sum_j v_{ij} c_j + \frac{1}{2}\sum_{j,k} v_{ijk} c_j c_k + \frac{1}{6}\sum_{j,k,l} v_{ijkl} c_j c_k c_l \quad (7)$$

$$M_{ij} = \gamma \sqrt{c_i(1-c_i)c_j(1-c_j)} \exp\frac{(u_i + u_j)}{2T} \quad (8)$$

where $\gamma = \langle w^{eff} \exp(-\hat{E}^{SP}/T) \rangle$ is a configuration-independent factor [16], and $u_i$ is a so-called "asymmetric" potential hereafter assumed to be equal to zero. The use of more accurate approximations than MFA (for example, the kinetic cluster method [16]) preserves the form of the master equation (5), but now with functions $\lambda_i$ and $M_{ij}$, given by more complex expressions. Here we restrict ourselves to the MFA, which allows us to reliably reproduce the main features of transformation kinetics as discussed in Ref. [17].

As has been discussed above, the lattice strain caused by the solute atoms and the resulting elastic stresses play a decisive role in the formation of GPZ and their subsequent transformation in Al-Cu alloys. The MSCE model proposed in Refs. [9,10,11], uses a mixed-space representation alloy configurational energy at two extreme length scales (atomic and continuum, respectively). Here, in order to build a more consistent model of GPZ formation, we use the approach of Ref. [14] in which the strain energy contribution is included directly into the effective cluster interactions. Since the lattice relaxation happens much faster than the growth of precipitates, for any alloy configuration given by the configuration numbers $n_i$, the strain field will almost instantaneously attain the relaxed values. Therefore, the effective interactions incorporating strain energy effects should be capable of describing the early stages of precipitation in the Al-Cu system.

It has been shown in calculations [14] that the strain contribution to the effective cluster interactions is predominant in Al-Cu alloys and that the precipitate morphology is determined by the three-body ($v_{ijk}^{(3)}$) and four-body ($v_{ijkl}^{(4)}$) cluster interactions. The calculated interaction energies for the most compact clusters on the fcc lattice are listed in Table 1. The negative sign of the pair interaction energy at the nearest-neighbor distance $v_1^{(2)}$ indicates the clustering tendency of Cu solutes in the Al matrix. Among the three- and four-body clusters of Cu solutes the energetically favored ones are isosceles right triangles denoted as (112) and squares denoted as (111122) belonging to {001} lattice plane, see Figs. 1 b, e.

To investigate the kinetics of solid solution decomposition and precipitate formation we have chosen the model interaction parameters of Hamiltonian Eq. (1) so that their ratios are close to those between the PAW-VASP calculated interactions [14]. Since the interaction parameters enter Eqs. (5), (7), (8) divided by temperature, we choose the value of $|v_1^{(2)}|$ to be the unit of energy and temperature in the present model.

Table 1. Effective cluster interaction energies calculated *ab initio* in Ref. [14] and used in this work. Cluster index (second column) lists the distances between all pairs of sites in the cluster.

|  | Cluster index | PAW-VASP [14] (eV) | This. ($|v_1^{(2)}|$) |
|---|---|---|---|
| $v_1^{(2)}$ | 1 | -0.04 | -1 |
| $v_2^{(2)}$ | 2 | -0.005 | 0 |
| $v_1^{(3)}$ | 111 | 0.037 | 1 |
| $v_2^{(3)}$ | 112 | -0.031 | -1 |
| $v_1^{(4)}$ | 111111 | 0.132 | 3 |
| $v_2^{(4)}$ | 111112 | 0.038 | 1 |
| $v_3^{(4)}$ | 111122 | -0.037 | -1 |

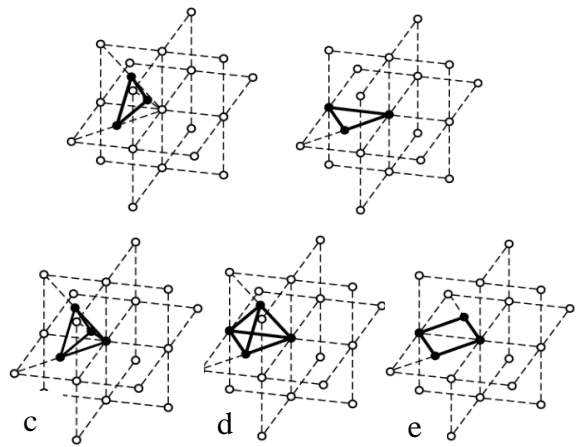

Figure 1. Three-body (a,b) and four-body (c,d,e) clusters corresponding to interactions $v_1^{(3)}$, $v_2^{(3)}$, $v_1^{(4)}$, $v_2^{(4)}$, and $v_3^{(4)}$ taken into account by the present model.

The phase diagram for the present model has been calculated using the standard thermodynamic relationships [8] in the mean-field approximation (MFA) which yields the following expression for the energy of quasi-equilibrium state of the alloy characterized by the set of local concentrations $c_i$:

$$F = T\sum_i \left(c_i \ln c_i + (1-c_i)\ln(1-c_i)\right) + H(c_i) \quad (9)$$

Here $H(c_i)$ represents a thermodynamic energy of quasi-equilibrium state of the alloy as given by Eq. (1), in which the occupation numbers $n_i$ have been replaced with their average values $c_i$.

In addition, we take into account the fact that the model with the chosen set of parameters predicts the formation of an ordered $L1_2$ type structure. Although the $AlCu_3$ phase is not known to form in real Al-Cu alloys, this result of the present model (restricted to the fcc underlying lattice and a very limited set of interactions) correctly captures the experimentally observed tendency in the Al-Cu system to form Cu-rich ordered compounds. In what follows, the $L1_2$ $AlCu_3$ phase will be considered as representing the whole manifold of ordered Cu-rich phases in the Al-Cu system.

Kinetic equations (5), (7), (8) were solved numerically for a cubic 40×40×40 simulation box on the fcc lattice subject to periodic boundary conditions. Since in the absence of fluctuations a uniform state of the system is a stationary solution, a perturbation corresponding to a Cu precipitate embryo was introduced in the initial concentration field $c_i(0)$. The embryo had a flat shape and a size ranging from 4×4×1 to 10×10×2 lattice parameters and was embedded in the otherwise uniformly disordered medium. The temporal evolution of the precipitate morphology, obtained as the solution of the kinetic equations was then monitored.

## 3. CALCULATION RESULTS

The phase diagram of Al-Cu alloys calculated using the free energy of Eq. (9) is presented in Fig. 2. The two lines labeled as 1 correspond to the boundary of stability solid solution with respect decomposition (binodal). This boundary is hereafter referred to as the decomposition line. Solid line 2 corresponds to the equilibrium between the disordered solid solution phase and the $L1_2$-ordered phase $AlCu_3$ with the same average atomic fraction of Cu atoms (i.e. without decomposition). This line is hereafter referred as the ordering line. In the region between the decomposition line on the left hand-side, and the ordering lines, a disordered solid solution is stable with respect to $L1_2$ ordering but unstable with respect to the formation of Cu-rich precipitates. At temperatures $T > 1.5$ (in units of $|v_1^{(2)}|$) and below the decomposition line on the left hand-side a two-phase state (solid solution plus Cu-rich precipitates) may be kinetically favored.

The results of the kinetic modeling by numerically solving Eqs. (5), (7), and (8) are presented in Figs. 3, 4, 5, and 6 for the values $c$ and $T$ variables, indicated by points labeled by A, B, C, and D in Fig. 2. The intensity of the gray color in Figs. 3-6 is proportional to the local Cu concentration $c_i$. The sites for which the $c_i$ is less than 0.5 are not shown in order not to obscure the other sites. The snapshots shown in Figs. 3-6 correspond to the moments of reduced time $t' = \gamma t$ specified in the figure captions.

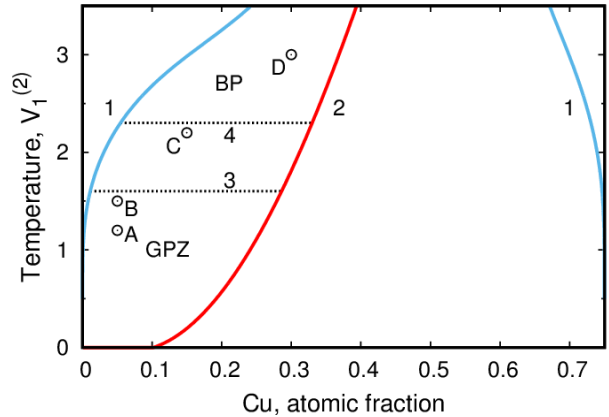

Figure 2. Calculated temperature–concentration phase diagram for the considered model of Al-Cu alloys. Solid lines labeled as 1 describe the stability limit of the solid solution, line 2 corresponds to the phase equilibrium between the disordered solid solution and the ordered $L1_2$ phase with the same average concentration of Cu atoms. Horizontal lines 3 and 4 separate the regions corresponding to different morphologies of Cu-rich precipitates: Label GPZ – designates the domain of platelet-like Guinie-Preston zones and BP – the domain of equiaxially shaped "bulky" precipitates). The dots labeled by the letters A, B, C, D show the values of the variables $c$ and $T$ that have been used in the simulations to study the precipitation kinetics.

For relatively low values of temperature and concentration (point A in the GPZ domain in Fig. 2) the initially introduced embryo extends along the (001) plane to form a single-plate Guinier-Preston zone (GP1). The evolution in the precipitate morphology observed in this case is shown in Fig. 3. The figure shows that along with the extension in the (001) plane, the plate becomes thicker in its central part to form perpendicular plates at the late stages (Figs. 3 c,d).

As mentioned in the previous Section, the clustering of Cu solutes in the {001} plane minimizes the alloy energy for the set of effective interactions $v_i^{(n)}$ given in Table 1. At low temperatures, the energy term in (9) dominates and favors the platelet morphology of the Cu precipitates.

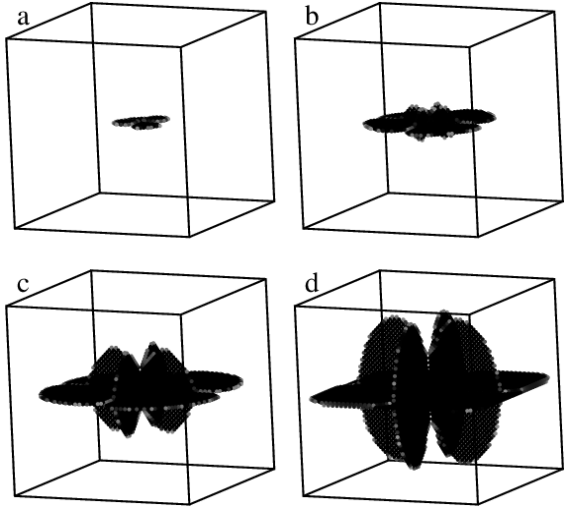

Figure 3. Temporal evolution of Cu solute distribution obtained from Eqs. (5), (7), (8) for values $c = 0.05$ and $T = 1.2$ (point A inside domain GPZ in Fig. 2). Snapshots a, b, c, and d were taken at times $t' = 10, 50, 100,$ and $150$, respectively.

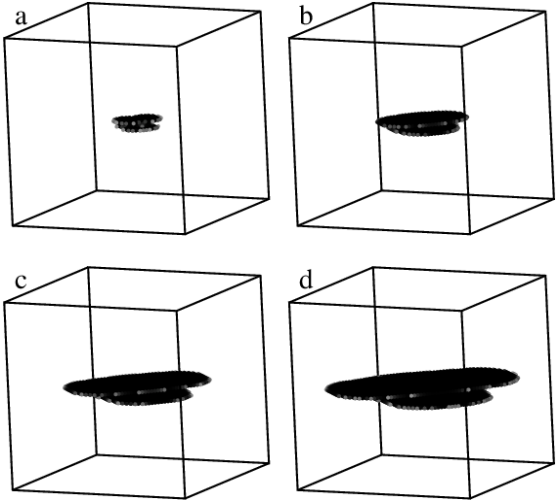

Figure 4. Similar to Fig. 3 but with $c = 0.05$ and $T = 1.5$ (point B inside the GPZ domain in Fig. 2). Snapshots a, b, c, and d were taken at time $t' = 10, 50, 100,$ and $150$, respectively.

With increasing temperature to values near the boundary of domain GPZ (point B in Fig. 2), the kinetics of solid solution decomposition change (Fig. 4). The precipitate that forms at these conditions has also the platelet morphology but it is comprised of two (001) planes of Cu atoms separated by a distance of 1.5 lattice parameters. Because the Cu atoms that belong to different (001) plates do not interact directly in the present model, the morphology of the double-layer precipitate observed in Fig. 4 (and similar to the morphology of GP2 zones) must have a kinetic origin. A detailed analysis of the precipitate formation in this case revealed that the precipitate thickness increases at early stages. After reaching a certain size, the precipitate became separated due to the displacement of atoms from the central layers to the two parallel (001) marginal plates of the precipitate. After which, these two Cu plates started to grow independently.

Inside a transient region corresponding to the temperature interval $1.6 < T < 2.3$, platelet-like and equiaxial precipitates are similar in terms of their free energies. In this region, the shape of the precipitate is strongly dependent on the initial conditions corresponding to the embryo. Fig. 5 shows that initially the precipitate had an equiaxial "bulky" morphology (snapshots a, b, and c). After a long period of time and under favorable conditions, the growth along the (001) plane occurred. Thus, two different precipitate morphologies can co-exist in a certain interval of holding time. As a result, the transition from the platelet to the equiaxial precipitate morphology is rather gradual, so that the boundaries indicated by lines 3 and 4 in the phase diagram Fig. 2 are indistinct. It should be mentioned that the evolution observed in Fig. 5 is qualitatively like the one shown in Fig. 4 but the difference is that in the former case the equiaxial shape is preserved for much longer holding times. In turn, the formation of GP zones inside the GPZ domain is almost independent of the shape of initial perturbation.

At high temperatures ($T > 2.3$, domain BP in Fig. 2) the entropy part of the free energy (first term in Eq. (9)) becomes dominant. In this case, the platelet morphology does not correspond to the free energy minimum and the solid solution decomposition yields precipitates with equiaxial morphology (Fig. 6). A small flat embryo created at the beginning becomes cube-shaped in a short period of time and subsequently grows, preserving the symmetry of the cubic lattice. The high-temperature domain, in which the growth occurs according to the above scenario (without the formation of platelet-like zones), is denoted in Fig. 2 as BP (bulky precipitates) and is bounded by the horizontal dotted line 4.

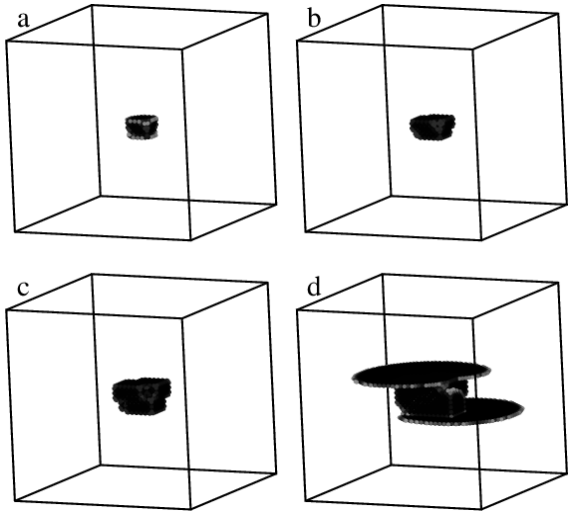

Figure 5. Similar to Fig. 3 but with $c = 0.15$ and $T = 2.2$ (point C inside the transient region in Fig. 2). Snapshots a, b, c, and d were taken at times $t' = 50, 100, 200$, and $400$, respectively.

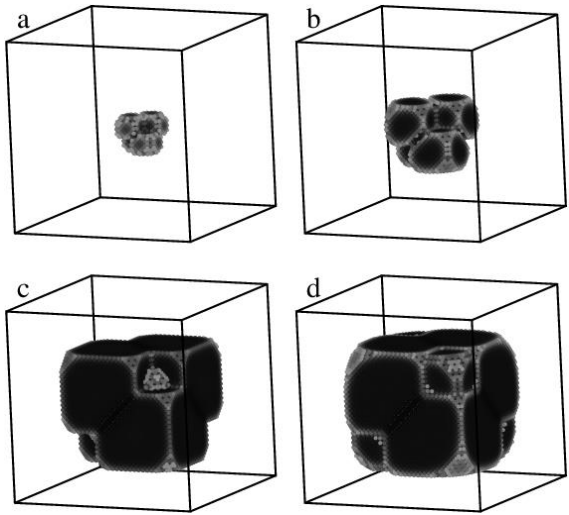

Figure 6. Similar to Fig. 3 but with $c = 0.3$ and $T = 3$ (point D inside domain BP in Fig. 2). Snapshots a, b, c, and d were taken at times $t' = 10, 50, 200, 1000$, respectively.

Our analysis of the distribution of Cu atoms on the fcc lattice inside the precipitates that form in the BP domain has shown that their structure corresponds to that of the ordered $Cu_3Al$ phase (structure type $L1_2$). Therefore, although the entropy increase with temperature prevents the formation of GPZ in the BP domain, this increase is insufficient for destroying the chemical order.

The dissolution of $Cu_3Al$ precipitates occurs with increase in temperature up to the binodal (line 1 on Fig. 2), above which the alloy is stable in the solid solution phase.

## 4. DISCUSSION

The Al-Cu alloy system is of practical relevance and represents a vivid example of a system characterized by anisotropic cluster interactions in which the strain contribution plays a determinant role [9,13,14]. Using statistical kinetic theory methods [16] with first-principles parametrization of the effective cluster interactions, we investigated in detail the kinetics of the early stages of the solid solution decomposition in the Al-Cu system as a function of temperature and concentration of Cu.

The kinetic master equation approach employed in our modeling considers the elastic strain by including it via many-body interactions. As a result, this approach is internally consistent, since it uses only the microscopic interaction model and does not include contributions from different length scales as it usually the case in phase field modeling. So, the approximations used here are more reliable for the description of early stages of precipitation. For later stages of the precipitate evolution, when the size of precipitates becomes large enough, the long-range elastic fields begin to play a significant role and the MSCE approach [9] becomes suitable.

We show that the formation of Guinier-Preston zones takes place at sufficiently low temperatures and may include several stages. Kinetic factors are found to play a significant role in determining the morphology of precipitates formed. In particular, a characteristic feature of the kinetics of the solid solution decomposition of Al-Cu is the formation of double-layer Guinier-Preston zones (Fig. 4). This process goes through a stage of increasing thickness of the initially one-platelet Cu precipitate with a subsequent dissolution of the inner layers of this precipitate with formation of two platelets in a parallel {001} plane. We believe that such a mechanism may be responsible for the formation of GP2 ($\theta''$–phase) precipitates.

Indeed, a similar multi-step process of GPZ formation was recently observed in simulations with Monte-Carlo – molecular dynamic approach [13]. To obtain the correct separation distance between the Cu plates in the GP2 zones observed experimentally [5], it is necessary to take into account longer-range interactions which incorporate the effects of coherency strain fields accompanying the formation of GP2 zones.

With the model considered here, the mechanism of solid solution decomposition is found to vary in a predictable way with both increasing temperature and alloy concentration (Section 3). For low concentrations of Cu that are typical of Cu-bearing Al alloys, an increase in temperature above the binodal line results in the dissolution of precipitates. Our model predicts the existence of a two-phase region where the precipitates are in equilibrium with the solid solution.

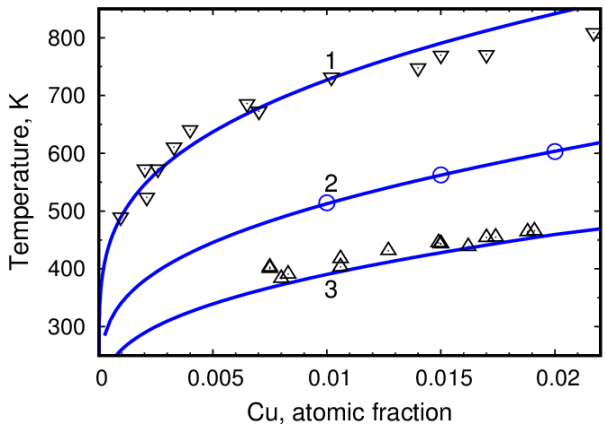

Figure 7. Calculated binodal line (1) and metastable GPZ solvus line (3) for low concentration Al–Cu alloys in comparison with experiment. Symbols ∇ and Δ mark experimental data for the binodal [21] and GP1 solvus [22], respectively. Line 2 represents the result of Monte-Carlo modeling [14] of solid solution decomposition to form GPZ.

In order to relate the results of the present modeling to experiment, in Fig. 7 we make a transformation from the dimensionless units of temperature used so far (see Fig. 2) to Kelvin by assigning to $v_1^{(2)}$ its value calculated in Ref. [14] (see Table 1). Fig. 7 shows that our simple model yields the binodal (line 1) in very good agreement with experiment. Such an agreement is unexpected and probably appears due to cancellation of errors caused by the neglect of vibrational entropy contribution [23] and considering AlCu$_3$ precipitates *in lieu* of $\theta$-phase.

The metastable solvus (line 3 in Fig. 7) is obtained by first choosing a temperature $T_1$ in between points A and B in Fig. 2 (corresponding, respectively, to GP1 and GP2 zone morphology in a 5 at.% Cu alloy) and then extrapolating it down in concentration $c$ using the on-site approximation for configurational entropy (see Ref. [23]), $T = T_1 \ln(0.05)/\ln(c)$. As it seen in Fig. 7, our calculations predict the position of the GPZ solvus very close to experiment [22].

Also shown in Fig. 7 are the results of *ab initio* based Monte Carlo simulations of GPZ formation in Al-Cu alloys in Ref. [14] (line 2). Although a more sophisticated interactions model is used in Ref. [14], the calculated temperatures of solid solution decomposition are overestimated compared to experiment [22] and are reaching the CALPHAD-derived iso-structural miscibility gap for Al–Cu fcc solid solution [19]. Compared to the present modeling, the region above line 2 is already in the domain indicated as BP in Fig.2 where the solid solution decomposition yields precipitates with equiaxial morphology.

The effective interactions used in Ref. [14] are temperature-independent and have been obtained from static ion calculations for supercells with various solute configurations in dilute Al–Cu alloys considered at the room-temperature calculated average lattice constant. In Ref. [23], the vibrational contribution to free energy has been shown to be responsible for a substantial increase of the solubility in the Al-Cu system. In Appendix we present the results of our own analysis of the vibrational contributions to the Cu solution energy in Al. At variance with Ref. [23], we employ a *quasiharmonic* description (accounting for thermal expansion) that yields similar values for the impurity solution entropy at moderate temperatures (100-500 K) as obtained in the harmonic approximation of Ref. [23] (see Appendix and Fig. A1 therein). We conclude that the deviation of line 2 in Fig. 7 from the experimental data on GPZ solvus is reasonable because the account of lattice vibrations would stabilize the Al-Cu solid solution to suppress its decomposition temperature in the MC simulation. A more definite conclusion about the role of lattice vibrations in the GPZ formation in Al-Cu alloys could be drawn if anharmonic effects and all the relevant atomic configurations are fully taken into account in a generalized model.

Despite the simplicity and numerous approximations involved, the model proposed in this study gives a fairly good description of the thermodynamics and kinetics of precipitation in Al-Cu alloys. In the framework of the model considered here, an increase in temperature (and/or Cu concentrations) produces a complex evolution of the structural state of the alloys and a change in the precipitate morphology from platelet GPZ to equiaxial Cu$_3$Al precipitates of the L1$_2$ type. There is also a transient region in which the two precipitate morphologies coexist. However, the results of the present modeling differ from the experimentally observed sequence GP1 → θ″ (GP2) → θ′ → θ (Al$_2$Cu) of phase transformations in Al—Cu alloys. The difference is due to simplifications employed in the present model as it describes the decomposition process on the underlying fcc lattice using a minimal set of interaction parameters. Additionally, the quenched-in vacancies may play a significant role in the solid solution decomposition [8]. Despite the above mentioned shortcomings, the obtained results of the model correctly reproduce the kinetics at the early stages of the decomposition of the solid solution at which the formation of GP1 and GP2 zones takes place.

This work has been supported by the Russian Foundation for Basic Research RFBR (grant 15-02-02084) and by Act 211 Government of the Russian Federation, contract № 02.A03.21.0006. P.A.K acknowledges support of the Swedish Foundation for Strategic Research (project RMA11-0090 "ALUX"). This work has been carried out using computing resources of the Federal center of collective usage for Complex Simulation and Data Processing for Mega-science Facilities at NRC "Kurchatov Institute", http://ckp.nrcki.ru/

**Appendix.**

The vibrational entropy of dilute solutions of Cu in Al has been evaluated in the quasiharmonic approximation using the full-potential projector augmented wave method [24] as implemented in the Vienna ab initio simulation package (VASP) [25]. The calculations were done for pure fcc Al, fcc Cu, and a 2×2×2(×4) Al-based fcc supercell containing a single Cu impurity. All atoms were allowed to relax during the calculations. The vibrational entropy of solution $\Delta S$ of Cu in Al is given by the total entropy of the $N$-atom supercell of the Al matrix with one Cu atom ($S_{Al_{N-1}Cu_1}$), taken with respect to the equivalent amounts of the Al ($S_{Al}$) and Cu ($S_{Cu}$) constituents, each in their fcc equilibrium crystal structures, and obtained as

$$\Delta S = S_{Al_{N-1}Cu_1} - \frac{N-1}{N} S_{Al_N} - \frac{1}{N} S_{Cu_N}.$$

The free energy of lattice vibrations was calculated in the quasiharmonic approximation using the PHONOPY code [26], where the dynamical matrix was obtained by the small displacement method [27]. The PBE generalized gradient approximation [28] was used. The self-consistent electronic structure calculations were done using a 14×14×14 k point mesh of Monkhorst and Pack (MP) [29]. The kinetic energy cutoff was 350 eV. The convergence tolerance for the total energy was $10^{-6}$ eV/atom and $10^{-3}$ eV/A for forces on atoms during local lattice relaxations. The phonon density of states and the vibrational free energies were evaluated using a uniform MP mesh of q points. The results of calculations are presented in Fig. A1.

It should be noted that Gibbs free energy of solution $\Delta G$ of Cu in Al at 0K and 300 K is equal, respectively, to -0.09 eV and -0.14 eV for the 32-atomic supercell. These values are in agreement with the data reported previously as -0.08 eV and -0.12 eV from LDA and GGA static-ion calculations for 64-atomic supercells [30] and -0.14 eV at 298 K from the COST507 CALPHAD database [31].

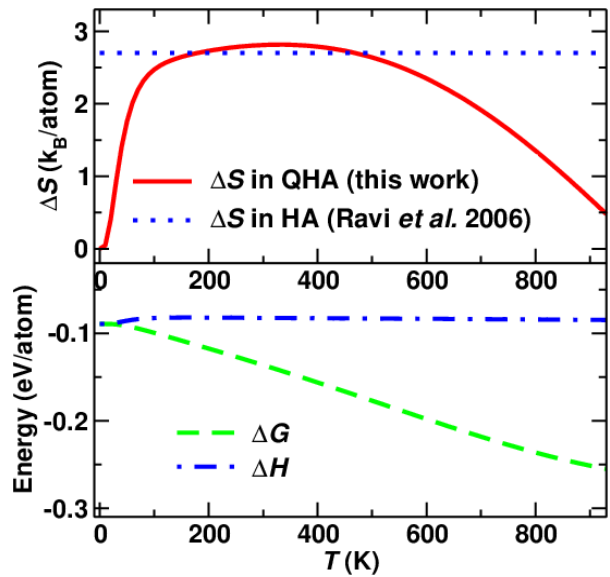

Figure A1. Top panel: Vibrational entropy of solution of a Cu impurity in Al calculated in the quasiharmonic approximation (solid line) compared with the result from Ref. [23] calculated in the harmonic approximation (dotted line). Bottom panel: Calculated Gibbs free energy $\Delta G$ and enthalpy $\Delta H$ of solution of Cu in Al.